\documentclass[aps,prd,preprint,superscriptaddress,amsmath,amssymb,showpacs]{revtex4-1}
\usepackage{dcolumn}
\usepackage{graphicx}
\usepackage{float}
\usepackage{physics}
\usepackage[colorlinks=true,allcolors=blue]{hyperref}

\begin{document}

\title{Potential analysis of holographic Schwinger effect in the magnetized background}

\author{Zhou-Run Zhu}
\email{zhuzhourun@mails.ccnu.edu.cn}
\affiliation{Institute of Particle Physics and Key Laboratory of Quark and Lepton Physics (MOS), Central China Normal University,
Wuhan 430079, China}

\author{De-fu Hou }
\thanks{Corresponding author}
\email{houdf@mail.ccnu.edu.cn}
\affiliation{Institute of Particle Physics and Key Laboratory of Quark and Lepton Physics (MOS), Central China Normal University,
Wuhan 430079, China}

\author{Xun Chen}
\email{chenxunhep@qq.com}
\affiliation{Institute of Particle Physics and Key Laboratory of Quark and Lepton Physics (MOS), Central China Normal University,
Wuhan 430079, China}

\date{\today}

\begin{abstract}

We study the holographic  Schwinger effect with  magnetic field at  RHIC and LHC energies by using the AdS/CFT correspondence. We consider both weak and strong  magnetic field cases with $B\ll T^2$  and $B\gg T^2$ solutions respectively. Firstly, we calculate separating length of the particle pairs at finite magnetic field. It is found that for both weak and strong magnetic field solutions the maximum value of separating length decreases with the increase of magnetic field , which can be inferred that the virtual electron-positron pairs become real particles more easily.  We also find that the magnetic field reduces the potential barrier and the critical field for the weak magnetic field solution, thus favors the Schwinger effect. With  strong magnetic field solution, the magnetic field enhances the Schwinger effect when the pairs are in perpendicular to the magnetic field although the magnetic field increases the critical electric  field.
\end{abstract}

\maketitle

\section{Introduction}\label{sec:01_intro}
 The virtual electron-positron pairs can be materialized under the strong electric-field in quantum electrodynamic (QED). This non-perturbative phenomenon is known as the Schwinger effect\cite{Schwinger:1951nm}. This phenomenon is not unique to QED, but has a general feature of vacuum instability in the presence of the external field. The production rate in the weak-coupling and weak-field case was put forward in \cite{Schwinger:1951nm} and was extend to the arbitrary-coupling and weak-field case\cite{Affleck:1981ag}:
 \begin{equation}
\label{eq1}
\ \Gamma\sim exp(-\frac{\pi m^{2}}{eE}+ \frac{e^{2}}{4}),
\end{equation}
where m, e represent the mass and charge of the particle pairs, respectively. E is the external electric-field. There exists a critical value $E_{c}$ of the electric field when the exponential suppression vanishes.

In string theory, there also exists a critical value $E_{c}$ which is proportional to the string tension \cite{Fradkin:1985ys,Bachas:1992bh}. By utilizing the AdS/CFT correspondence\cite{Maldacena:1997re,Gubser:1998bc,Witten:1998qj,Rey:1998bq}, the duality between the string theory on $AdS_{5}\times S_{5}$ space and the $\mathcal{N} = 4$ super Yang-Mills (SYM) theory, one can study the Schwinger effect in this holographic method. In order to realize the $\mathcal{N} = 4$ SYM system coupled with an U(1) gauge field, one can break the gauge group from $U(N+1)$ to $SU(N)\times U(1)$ by using the Higgs mechanism. In the usual studies, the test particles are assumed to be heavy quark limit. To avoid pair creation suppressed by the divergent mass, the location of the probe D3-brane is at finite radial position rather than at the AdS boundary. The mass of the particles is finite so that the production rate can make sense\cite{Semenoff:2011ng}. Therefore, the production rate can be given as
 \begin{equation}
\label{eq2}
\ \Gamma\sim exp[-\frac{\sqrt{\lambda}}{2}(\sqrt{\frac{E_{c}}{E}}-\sqrt{\frac{E}{E_{c}}})^{2}],
\end{equation}
with a critical field
 \begin{equation}
\label{eq3}
\ E_{c}=\frac{2\pi m^{2}}{\sqrt{\lambda}},
\end{equation}
which agrees with the result from the Dirac-Born-Infeld (DBI) action and $\lambda$ is the 't Hooft coupling.

Following the holographic step, the potential analysis was performed in the confining theories in \cite{Sato:2013dwa,Sato:2013iua}. The potential barrier can be regarded as a quantum tunneling process. The virtual particle pairs need to get enough energy from an external electric field. When reaching to a critical value $E_{c}$ the potential barrier will vanish. Then the real particles pairs production are completely uncontrolled and the vacuum turns into totally instability. The potential analysis provide a new perspective to study the Schwenger effect. A lot of research work have been studied by using the AdS/CFT correspondence. The production rate in the confining theories was discussed in \cite{Zhang:2015bha,Kawai:2013xya,Ghodrati:2015rta}. The universal nature of holographic Schwinger effect in general confining backgrounds was analyzed in \cite{Sato:2013hyw}. The Schwinger effect also has been investigated in the AdS/QCD models \cite{Hashimoto:2014yya,Sadeghi:2016ppy}. The potential analysis in non-relativistic backgrounds\cite{Fadafan:2015iwa} and a D-instantons background \cite{,Dehghani:2015gtd} were discussed. The holographic Schwinger effect in de Sitter space has been studied in \cite{Fischler:2014ama}. Other important research results can be seen in \cite{Zhang:2018hfd,Zhang:2018oie, Bolognesi:2012gr,Sato:2013pxa,Zhang:2017dmf,Zhang:2017egb,Dietrich:2014ala,Hashimoto:2014dza,Wu:2015krf,Ambjorn:2011wz,Chakrabortty:2014kma,Chakrabortty:2014kma,Chowdhury:2019mqi}.

The heavy ion collisions at RHIC and LHC experiments produce strong electro-magnetic fields. As a result, studying the Schwinger effect in the strong magnetic field ($m^2_\pi \sim 15m^2_\pi$) created by RHIC and LHC\cite{Voronyuk:2011jd,Bzdak:2011yy,Skokov:2009qp,Deng:2012pc,She:2017icp} is the main motivation of this paper. The strong magnetic fields may provide us some different views for the vacuum structure and we expect the Schwinger effect may be observed through the heavy-ion collisions experiments in future. The magnetic field is expected to remain large enough when QGP forms although rapidly decays after the collision\cite{Tuchin:2013apa,McLerran:2013hla}. It has significant implications for the QCD matter near the deconfinement transition temperature \cite{DElia:2010abb} and QCD phase structure \cite{Miransky:2015ava,Bali:2011qj}. This expectation led to an in-depth research of QCD in the magnetized background. The asymptotically magnetic brane solutions were constructed in \cite{DHoker:2009mmn,DHoker:2009ixq}  in the $AdS_5$ of the Einstein-Maxwell theory which is dual to the $\mathcal{N} = 4$ SYM theory. The chiral magnetic effect in \cite{Fukushima:2008xe,Kharzeev:2007jp} has been studied. (Inverse) magnetic catalysis can see \cite{Miransky:2002rp,Mamo:2015dea,Li:2016gfn,Fang:2016cnt,Bohra:2019ebj,Kashiwa:2011js,Bruckmann:2013oba,Fukushima:2012kc,Ferreira:2014kpa,Bali:2013esa} and the holographic energy loss in the magnetized background see \cite{Zhu:2019ujc}. The magnetic field also has an influence on the early universe physics\cite{Grasso:2000wj,Vachaspati:1991nm}.

Thence, we study the holographic Schwinger effect in the 5-dimensional Einstein-Maxwell system with a proper magnetic field range \cite{Mamo:2015dea} produced in the non-central heavy-ion collisions  at RHIC and LHC  energies. This may give us some inspiration for studying the Schwinger effect through the experimental results. The production rate of Schwinger effect with the presence of electric and magnetic fields was discussed in \cite{Sato:2013pxa}. One way to turn on magnetic fields is considering a circular Wilson loop under the parallel electric and magnetic fields. Another way is to utilize circular Wilson loop solutions depending on additional parameters which are related to the magnetic fields. However these methods of adding magnetic field neglected the magnetic effect on the geometry of background. In this paper we incorporate a magnetic field  with the  magnetized Einstein-Maxwell system.
With the magnetized background in this paper , we study the holographice Schwinger effect with a magnetic field by using the AdS/CFT correspondence . The organization of the paper is as follows. In Sec.~\ref{sec:02}, we introduce  the 5-dimensional  Einstein-Maxwell system with a magnetic field. In Sec.~\ref{sec:03}, we study the potential analysis in the magnetized background with $B\ll T^2$ solutions. In Sec.~\ref{sec:04}, we discuss the the potential analysis when $B\gg T^2$. The discussion and conclusion are given in Sec.~\ref{sec:05}.

\section{Background geometry}\label{sec:02}
The gravity background with magnetic field was introduced into the 5-dimensional  Einstein-Maxwell system by using the AdS/QCD model \cite{DHoker:2009ixq}, and the action is

\begin{equation}
\label{eq4}
\ S = \frac{1}{16\pi G_5 }\int \dd{x}^5{ \sqrt{- g}(R-F^{MN}F_{MN}+\frac{12}{L^{2} })},
\end{equation}
where $g$ is the determinant of metric $g_{MN}$. $R$, $G_{5}$, $F_{MN}$ are the scalar curvature, $5D$ Newton constant and the U(1) gauge field, respectively. $L$ is the AdS radius and we set it to $1$.

As discussed in \cite{Mamo:2015dea}, turning on a bulk magnetic field in the $x_{3}-$direction and the metric of the black hole takes the form
\begin{equation}
\label{eq5}
\ ds^{2}=r^{2}(-f(r)dt^{2}+h(r)(dx_{1}^{2}+dx_{2}^{2})+q(r)dx_{3}^{2})+\frac{dr^{2}}{r^{2} f(r)},
\end{equation}
with
\begin{equation}
\label{eq6}
  f(r) = 1- \frac{r_{h}^{4}}{r^{4}}+\frac{2B^{2}}{3r^{4}}\ln(\frac{r_{h}}{r}),
 \end{equation}
 \begin{equation}
 \label{eq7}
  h(r) = 1+ \frac{1}{3} B^{2}\frac{\ln(r)}{r^{4}},
\end{equation}
 \begin{equation}
 \label{eq8}
  q(r) = 1- \frac{2}{3} B^{2}\frac{\ln(r)}{r^{4}},
 \end{equation}

where $r$ denotes the radial coordinate of the $5$th dimension. The magnetic field breaks the rotation symmetry and allows us to analyze the anisotropic cases because the element q(r) is not equal to h(r) and the anisotropy was induced by the magnetic field \cite{Giataganas:2013hwa,Finazzo:2016mhm}. The anisotropic direction is along $x_{3}-$direction in this article. The perturbative solutions of this black hole metric can work well when $B\ll T^{2}$. Note that the physical magnetic field $\mathfrak{B}$ is related with the magnetic field $B$ by the equation $\mathfrak{B}=\sqrt{3}B$.

The Hawking temperature is
\begin{equation}
\label{eq9}
T=\frac{r_h}{\pi}-\frac{B^{2}}{6\pi r_{h}^{3}},
\end{equation}
where $r_{h}$ is the black-hole horizon. In this article, we will use this Einstein-Maxwell system and extend it to study the holographic effect of magnetic field on the Schwinger effect.

\section{Potential analysis with weak magnetic field  $B\ll T^2$ solutions}\label{sec:03}

Since the magnetic field is along $x_{3}-$direction, it is reasonable to consider the test particle pairs are transverse to the magnetic field and parallel to the magnetic field. From this point of view, we perform the potential analysis with the two cases in the magnetized background.
\subsection{Transverse to the magnetic field}
We study the potential analysis with the test particle pairs separated in the $x_{1}-$direction first, which means the particle pairs are transverse to the magnetic field. The coordinates are parameterized by
\begin{equation}
\label{eq10}
\ t=\tau,\quad x_{1}=\sigma,\quad  x_{2}=x_{3}=0,\quad r=r(\sigma).
\end{equation}

By utilizing the Euclidean signature, the Nambu-Goto action is given as
\begin{equation}
\label{eq11}
 S= T_{F}\int d \sigma d\tau \mathcal{L}= T_{F}\int d \sigma d\tau\sqrt{\det g_{\alpha\beta}},
\end{equation}
where $g_{\alpha\beta}$ represents the determinant of the induced metric. $T_{F}=\frac{1}{2\pi \alpha'}$ is the string tension and
\begin{equation}
\label{eq12}
\ g_{\alpha\beta} = g_{\mu\nu}\frac{\partial X^{\mu}}{\partial \sigma^{\alpha}}\frac{\partial X^{\nu}}{\partial \sigma^{\beta}},
\end{equation}
where $g_{\mu\nu}$ denote the brane metric and $X^{\mu}$ is target space coordinates.

Then the induced metric is
\begin{equation}
\label{eq13}
\ g_{00}=r^{2}f(r),\quad g_{11}=r^{2}h(r)+\frac{1}{r^{2}f(r)}\dot{r}^{2},\quad g_{10}=g_{01}=0,
\end{equation}
with $\dot{r}=\frac{dr}{d \sigma}$.

The Lagrangian density is given as
\begin{equation}
\begin{split}
\label{eq14}
\mathcal{L}= \sqrt{\det g_{\alpha\beta}}=\sqrt{r^{4}f(r)h(r)+\dot{r}^{2}},
\end{split}
\end{equation}
and $\mathcal{L}$ does not rely on $\sigma$ explicitly. The conserved quantity is obtained by
\begin{equation}
\label{eq15}
\ \mathcal{L}-\frac{\partial \mathcal{L}}{\partial{\dot{r}}}\dot{r}=C,
\end{equation}
which leads to
\begin{equation}
\label{eq16}
\ \frac{r^{4}f(r)h(r)}{\sqrt{r^{4}f(r)h(r)+\dot{r}^{2}}}=C.
\end{equation}

 By using the boundary condition
\begin{equation}
\label{eq17}
\ \frac{dr}{d \sigma}=0, \quad r=r_{c} \ (r_{h}<r_{c}<r_{0}),
\end{equation}
where the D3-brane located at finite radial position $r = r_{0}$. The conserved quantity C can be expressed as
\begin{equation}
\label{eq18}
\ C=r_{c}^{2}\sqrt{f(r_{c})h(r_{c})}.
\end{equation}

Plugging Eq.(\ref{eq18}) into Eq.(\ref{eq16}),one get
\begin{equation}
\label{eq19}
\ \dot{r}=\frac{dr}{d \sigma}=r^{2}\sqrt{h(r)f(r)[\frac{r^{4}h(r)f(r)}{r_{c}^{4}h(r_{c})f(r_{c})}-1]}.
\end{equation}

By integrating Eq.(\ref{eq19}), one can get the separate length $x_{\bot}$ of the test particle pairs
\begin{equation}
\label{eq20}
\ x_{\bot}=\frac{2}{a r_{0}} \int^{\frac{1}{a}}_{1} dy \frac{1}{y^{2}\sqrt{f(r)h(r)[y^{4} \frac{f(r)h(r)}{f(r_{c})h(r_{c})}-1]}},
\end{equation}
with the dimensionless parameter
\begin{equation}
\label{eq21}
\ y\equiv \frac{r}{r_{c}},\ \ a\equiv \frac{r_{c}}{r_{0}}.
\end{equation}

By using Eq.(\ref{eq14}) and Eq.(\ref{eq19}), the sum of the Coulomb potential and static energy can be given as
\begin{equation}
\label{eq22}
\begin{split}
\ V_{(CP+SE)(\perp)} & = 2T_{F} \int^{\frac{x_{\bot}}{2}}_{0} d\sigma \mathcal{L} \\
  &= 2T_{F}a r_{0}\int^{\frac{1}{a}}_{1} dy \frac{y^{2}\sqrt{f(r)h(r)}}{\sqrt{y^{4}f(r)h(r)-f(r_{c})h(r_{c})}}.
\end{split}
\end{equation}

The critical field is obtained by the DBI action in the Lorentzian signature. The DBI action is
\begin{equation}
\label{eq23}
\ S_{DBI}=-T_{D3} \int d^{4}x \sqrt{-det(G_{\mu\nu}+\mathcal{F}_{\mu\nu})} ,
\end{equation}
with a D3-brane tension
\begin{equation}
\label{eq24}
\ T_{D3}=\frac{1}{g_{s}(2\pi)^{3}\alpha'^{2}}.
\end{equation}

From Eq.(\ref{eq5}), the induced metric $G_{\mu\nu}$ reads
\begin{equation}
\label{eq25}
\ G_{00}=-r^{2}f(r),\quad G_{11}=G_{22}=r^{2}h(r),\quad G_{33}=r^{2}q(r).
\end{equation}

Then considering $\mathcal{F}_{\mu\nu}=2\pi \alpha' F_{\mu\nu}$ \cite{Zwiebach:2004tj} and the electric field $E$ is along $x_{1}-$direction\cite{Sato:2013iua}, one gets
\begin{equation}
\label{eq26}
G_{\mu\nu}+\mathcal{F}_{\mu\nu}=
\left(\begin{array}{cccc}
    -r^{2}f(r)     & 2\pi \alpha'E    & 0   & 0 \\
    -2\pi \alpha'E &   r^{2}h(r)     & 0    & 0\\
    0 &                0&        r^{2}h(r) & 0\\
     0    &0     & 0     & r^{2}q(r)
\end{array}\right),
\end{equation}
which leads to
\begin{equation}
\label{eq27}
\ det(G_{\mu\nu}+\mathcal{F}_{\mu\nu})=-r^{4}h(r)q(r)[r^{4}f(r)h(r)-(2\pi \alpha')^{2}E^{2}].
\end{equation}

By plugging Eq.(\ref{eq27}) into Eq.(\ref{eq23}), one get
\begin{equation}
\label{eq28}
\ S_{DBI}=-T_{D3} \int d^{4}x \sqrt{r^{4}_{0}h(r_{0})q(r_{0})} \sqrt{r^{4}_{0}f(r_{0})h(r_{0})-(2\pi \alpha')^{2}E^{2}}.
\end{equation}
where $r=r_{0}$ is the location of the D3-brane. To avoid Eq.(\ref{eq28}) being ill-defined,
\begin{equation}
\label{eq29}
\ r^{4}_{0}h(r_{0})f(r_{0})-(2\pi \alpha')^{2}E^{2}\geq0 .
\end{equation}

The critical field $E_{c}$ is obtained by
\begin{equation}
\label{eq30}
\ E_{c}=T_{F}r^{2}_{0} \sqrt{f(r_{0})h(r_{0})}.
\end{equation}

In Eq.(\ref{eq30}), one can see that the critical field is related to the magnetic field. By introducing a dimensionless parameter $\alpha \equiv \frac{E}{E_{c}}$, the total potential $V_{tot(\perp)}$ is
\begin{equation}
\label{eq31}
\begin{split}
\ V_{tot(\perp)}&=V_{(CP+SE)(\perp)}-Ex_{\perp}  \\
              &=2T_{F}a r_{0}\int^{\frac{1}{a}}_{1} dy \frac{y^{2}\sqrt{f(r)h(r)}}{\sqrt{y^{4}f(r)h(r)-f(r_{c})h(r_{c})}}  \\
        &-\frac{2T_{F}\alpha r_{0}}{a}\int^{\frac{1}{a}}_{1} dy\frac{\sqrt{f(r_{0})h(r_{0})}\sqrt{f(r_{c})h(r_{c})}}{y^{2}\sqrt{f(r)h(r)[y^{4}f(r)h(r)-f(r_{c})h(r_{c})]}}.
  \end{split}
\end{equation}

\subsection{Parallel to the magnetic field}
We consider the test particle pairs separated in the $x_{3}-$direction which means the particle pairs are parallel to the magnetic field. The coordinates are parameterized by
\begin{equation}
\label{eq32}
\ t=\tau,\quad x_{3}=\sigma,\quad  x_{1}=x_{2}=0,\quad r=r(\sigma).
\end{equation}

By repeating the previous calculation, one can get the separate length $x_{\|}$
\begin{equation}
\label{eq33}
\ x_{\|}=\frac{2}{a r_{0}} \int^{\frac{1}{a}}_{1} dy \frac{1}{y^{2}\sqrt{f(r)q(r)[y^{4} \frac{f(r)q(r)}{f(r_{c})q(r_{c})}-1]}}.
\end{equation}

\begin{figure}
    \centering
      \setlength{\abovecaptionskip}{-0.1cm}
    \includegraphics[width=7.5cm]{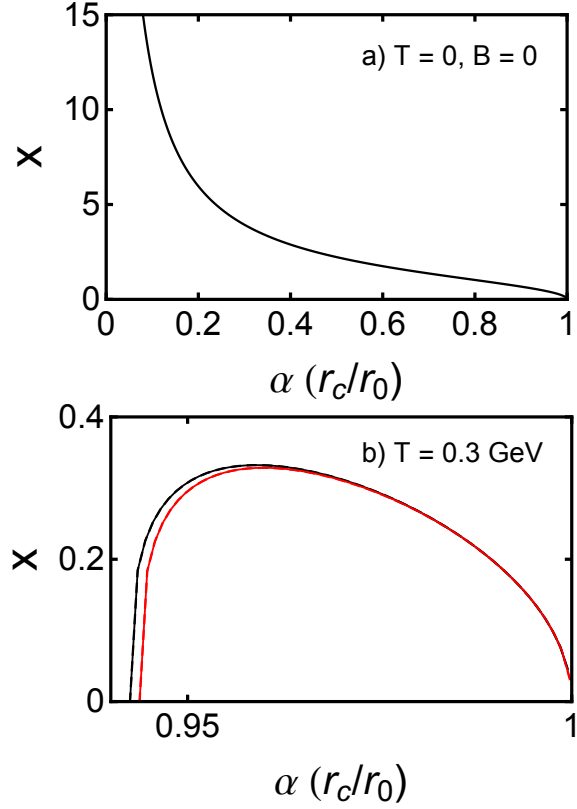}
    \caption{\label{fig1} The separate length x versus the parameter $a(r_c/r_0)$. a) for $T = 0,\ B = 0$, (b) for $T = 0.3\ GeV$. The black line and red line in b) denote $B = 0.01\ GeV^{2},\ 0.08\ GeV^{2}$, respectively. The solid line in b) indicates the particle pair is parallel to the magnetic field direction, and the dashed line is perpendicular to the magnetic field direction.}
\end{figure}

\begin{figure}
    \centering
      \setlength{\abovecaptionskip}{-0.1cm}
    \includegraphics[width=7.5cm]{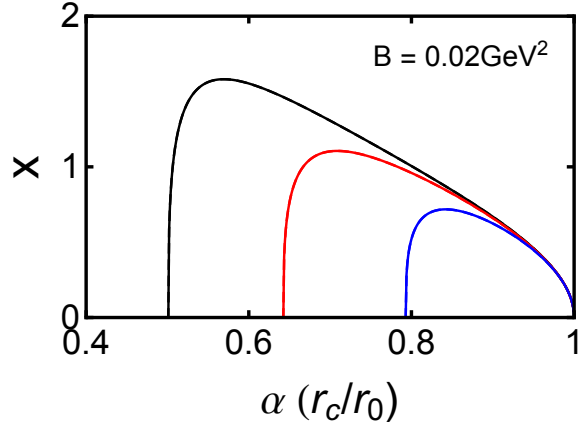}
    \caption{\label{fig2} The separate length x versus the parameter $a(r_c/r_0)$ for different temperature when $B = 0.02\ GeV^{2}$. The black line, red line, blue line denote $T = 0.2\ GeV,\ 0.25\ GeV,\ 0.3\ GeV$, respectively. The solid line indicates the particle pair is parallel to the magnetic field, and the dashed line is perpendicular to the magnetic field.}
\end{figure}
The separate length x  versus the parameter $a=r_c/r_0$ in different situations is depicted  in Fig.~\ref{fig1} and Fig.~\ref{fig2}. First, we note that there are two possible U-shape string configurations, samilar as heavy quark limit\cite{Rey:1998bq,Colangelo:2010pe,Chen:2017lsf}. The U-shape string  remains unchanged at vanishing temperature for all separate distance, while the U-Shape string exists only at large $a$ and become unstable at small $a$ for finite temperature case. We take the stable branch, corresponding to large values of $a$ in the potential analysis. In our numerical computation, we set $T_{F}$ and $r_{0}$ as constants for simplicity. Next, from these two pictures, we can see that the maximum value of distance is decreasing with the increases of temperature and magnetic field. Thus we can infer that Schwinger effect happens easily at larger temperature and magnetic field.

The sum of the Coulomb potential and static energy at the finite temperature in the magnetized background is
\begin{equation}
\label{eq34}
\ V_{(CP+SE)(\|)}= 2T_{F}a r_{0}\int^{\frac{1}{a}}_{1} dy \frac{y^{2}\sqrt{f(r)q(r)}}{\sqrt{y^{4}f(r)q(r)-f(r_{c})q(r_{c})}}.
\end{equation}

The the total potential $V_{tot(\|)}$ can be obtained as
\begin{equation}
\label{eq35}
\begin{split}
        V_{tot(\|)}&=V_{(CP+SE)(\|)}-Ex_{\|}  \\
        &=2T_{F}a r_{0}\int^{\frac{1}{a}}_{1} dy \frac{y^{2}\sqrt{f(r)q(r)}}{\sqrt{y^{4}f(r)q(r)-f(r_{c})q(r_{c})}}  \\
        &-\frac{2T_{F}\alpha r_{0}}{a}\int^{\frac{1}{a}}_{1} dy\frac{\sqrt{f(r_{0})h(r_{0})}\sqrt{f(r_{c})q(r_{c})}}{y^{2}\sqrt{f(r)q(r)[y^{4}f(r)q(r)-f(r_{c})q(r_{c})]}}.
  \end{split}
\end{equation}

\begin{figure}[H]
    \centering
      \setlength{\abovecaptionskip}{-0.1cm}
    \includegraphics[width=7.5cm]{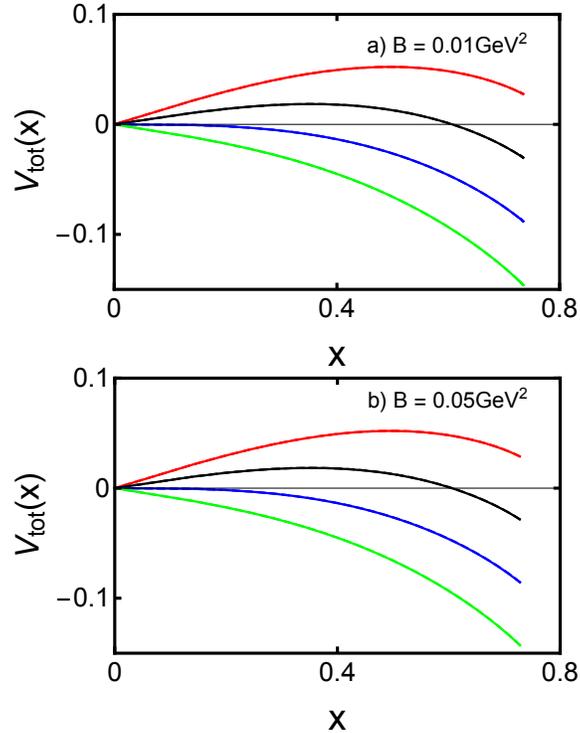}
    \caption{\label{fig3} The total potential $V_{tot}$ with respect to the separate length x with different electric field when $T = 0.25\ GeV$. The red line, black line, blue line, green line denote  $\alpha = 0.8,\ 0.9,\ 1.0,\ 1.1$, respectively. a) for $B = 0.01\ GeV^{2}$ and b) for $B = 0.05\ GeV^{2}$. The solid line (dashed line) indicates the particle pair is parallel (perpendicular) to the magnetic field.}
\end{figure}

The shapes of the total potential $V_{tot}$ with respect to the separate length x for various $\alpha$ when $T = 0.25\ GeV$ are plotted in Fig.~\ref{fig3}. We can find that the potential barrier decreases with the increase of external electric-field and  vanishes at a critical field. When $\alpha <1$, the potential barrier is existent and the pairs production can be explained by the tunneling process. When $\alpha >1$, the particles are easier to produce as the external electric-field increases. The vacuum becomes unstable extremely and the production of the pairs are explosive. The result agrees with the shapes of the potential for various values of $E_{c}$ in \cite{Sato:2013iua}.

\begin{figure}[H]
    \centering
    \setlength{\abovecaptionskip}{-0.1cm}
    \includegraphics[width=7.5cm]{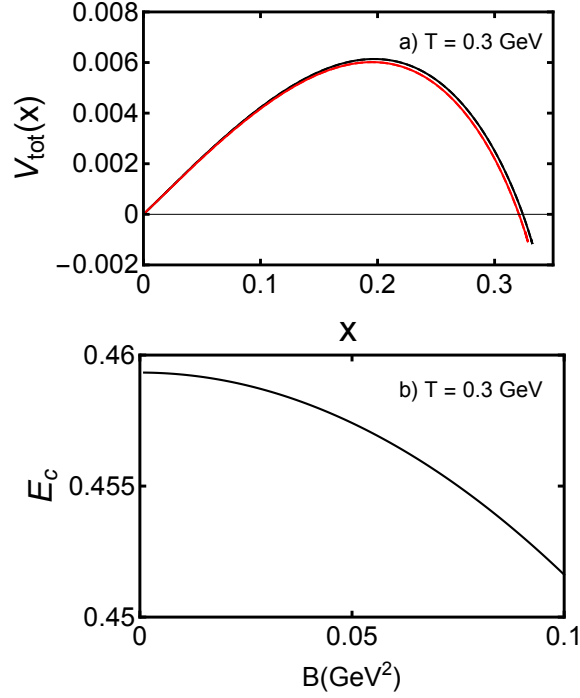}
    \caption{\label{fig4} a) for the total potential $V_{tot}$ against the separate length x with $\alpha = 0.9$ for the different magnetic fields when $T = 0.3\ GeV$. The black line and red line in a) denote $B = 0.01\ GeV^{2}$, $0.08\ GeV^{2}$, respectively. The solid line (dashed line) indicates the particle pair is parallel (perpendicular) to the magnetic field. b) for $E_{c}$ against $B$ when $T = 0.3\ GeV$.}
\end{figure}

The effect of the magnetic field on the total potential when $T = 0.3\ GeV$ is studied in Fig.~\ref{fig4}. We find that the magnetic field reduces the height and width of the potential barrier and favor the Schwinger effect in $a)$. We also plot $E_{c}$ versus $B$ in $b)$. One can obtain that $E_c$ decreases as the magnetic field increases, so that Schwinger effect occurs easily. This result agrees with the finding of $a)$. The Schwinger effect is  more obvious when pairs are perpendicular to the magnetic field than that in parallel case .

\begin{figure}[H]
    \centering
      \setlength{\abovecaptionskip}{-0.1cm}
    \includegraphics[width=7.5cm]{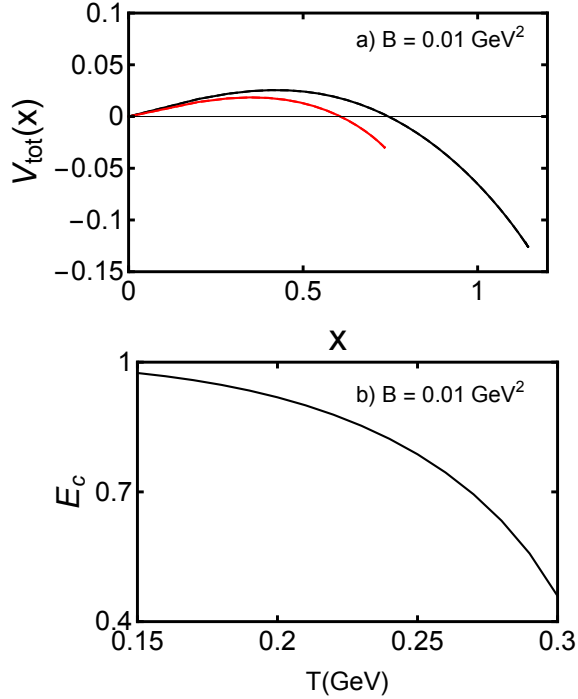}
    \caption{\label{fig5} a) for the total potential $V_{tot}$ against the separate length x with $\alpha = 0.9$ for the different T when $B = 0.01\ GeV^{2}$. The black line (red line) in a) denotes $T = 0.2\ GeV$ ($T = 0.25\ GeV$). The solid line (dashed line) indicates the particle pair is parallel (perpendicular) to the magnetic field. b) for $E_{c}$ against $T$ when $B = 0.01\ GeV^{2}$.}
\end{figure}

The relationship between the total potential and the temperature when $B = 0.01\ GeV^{2}$ is analyzed in Fig.~\ref{fig5}. One can see that the potential barrier decreases with the incrtease of temperature  in $a)$. It is found that the temperature also reduces the critical electric field $E_{c}$ in $b)$ and thus favors  the Schwinger effect.

\section{Potential analysis with strong magnetic field $B\gg T^2$ solutions}\label{sec:04}
In this section, we discuss the Schwinger effect for strong magnetic field case with $B\gg T^2$. In \cite{DHoker:2009mmn}, the $BTZ \times T^2$ black hole solution when $B\gg T^2$ is obtained
\begin{equation}
\label{eq36}
\ ds^{2}=3r^{2}(-f(r)dt^{2}+dx_{3}^{2})+ \frac{B}{\sqrt{3}}(dx_{1}^{2}+dx_{2}^{2})+ \frac{dr^{2}}{3r^{2} f(r)},
\end{equation}
with
\begin{equation}
\label{eq37}
  f(r) = 1- \frac{r_{h}^{2}}{r^{2}}.
 \end{equation}
The magnetic field is in $x_{3}-$direction in this black hole. The Hawking temperature is
\begin{equation}
 \label{eq38}
 T =  \frac{3 r_h}{2 \pi}.
 \end{equation}

When the particle pairs separated in the $x_{1}-$direction which means pairs are perpendicular to the magnetic field. The electric field $E$ is along $x_{1}-$direction, then the critical field $E_{c}$ and total potential $V_{tot}$ are
\begin{equation}
 \label{eq39}
 E_c = T_{F}r_{0} \sqrt{\sqrt{3} f(r_{0}) B},
 \end{equation}

\begin{equation}
\label{eq40}
\begin{split}
        V_{tot}&= 2 T_{F} a r_{0}\int^{\frac{1}{a}}_{1} dy \frac{\sqrt{A(r)}}{\sqrt{A(r)-A(r_c)}}-2 T_{F} a \alpha r^2_{0} \sqrt{\sqrt{3} f(r_{0}) B} \int^{\frac{1}{a}}_{1}dy \frac{\sqrt{A(r_c)}}{\sqrt{A^2(r)-A(r)A(r_c)}},
  \end{split}
\end{equation}
where
\begin{equation}
 \label{eq40}
 A(r) =\sqrt{3} r^2 f(r) B,\ A(r_c) =\sqrt{3} r^2_c f(r_c) B.
 \end{equation}

\begin{figure}[H]
    \centering
      \setlength{\abovecaptionskip}{-0.1cm}
    \includegraphics[width=7.5cm]{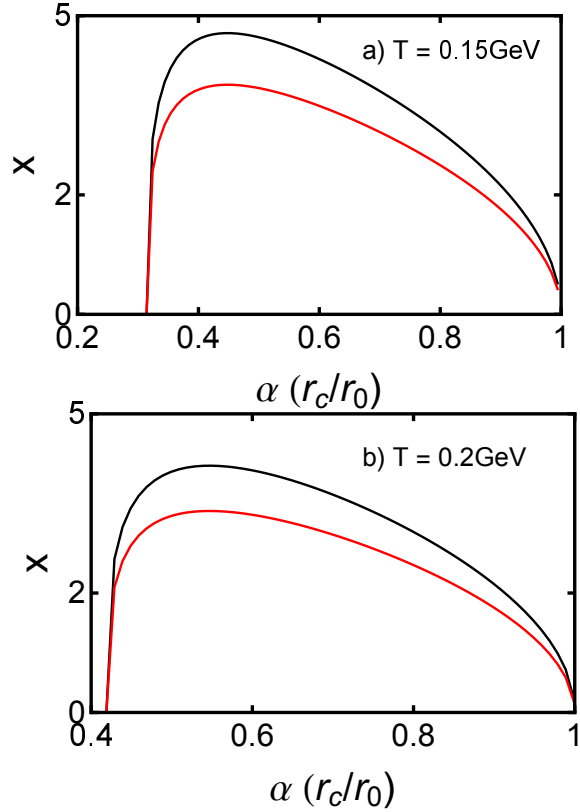}
    \caption{\label{fig6} The separate length x versus the parameter $a(r_c/r_0)$ in different temperature. a) for $T =0.15\ GeV$, b) for $T = 0.2\ GeV$. The black line and red line in a) and b) denote $B = 0.1\ GeV^{2},\ 0.15\ GeV^{2}$, respectively.}
\end{figure}

\begin{figure}[H]
    \centering
      \setlength{\abovecaptionskip}{-0.1cm}
    \includegraphics[width=7cm]{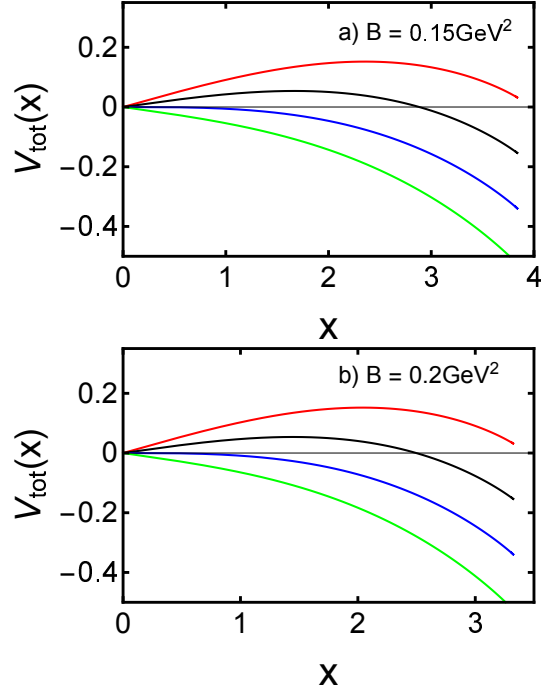}
    \caption{\label{fig7} The total potential $V_{tot}$ with respect to the separate length x with different electric field when $T = 0.15\ GeV$. The red line, black line, blue line, green line denote $\alpha = 0.8,\ 0.9,\ 1.0,\ 1.1$, respectively. a) for $B = 0.15\ GeV^{2}$ and b) for $B = 0.2\ GeV^{2}$.}
\end{figure}

The separate length x versus the parameter $a$ in different situations are plotted in Fig.~\ref{fig6}. We can find that the maximum value of distance is decreasing with the increasing magnetic field which is consistent with the results of Fig.~\ref{fig1}. The shapes of the total potential $V_{tot}$ versus the separate length x for various $\alpha$ when $T = 0.15\ GeV$ are plotted in Fig.~\ref{fig7}. When $\alpha <1$, the Schwinger effect can not occur. The potential barrier decreases with the external electric-field increasing. When $\alpha \geq 1$, the production of the pairs is not limited.

\begin{figure}[H]
    \centering
    \setlength{\abovecaptionskip}{-0.1cm}
    \includegraphics[width=7cm]{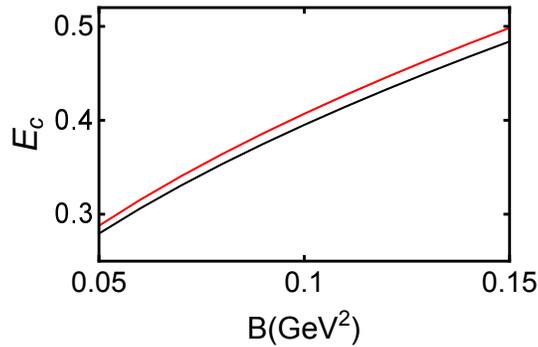}
     \caption{\label{fig8} $E_{c}$ against $B$ in different temperature. The red line and black line denote $T = 0.1\ GeV$ and $T = 0.15\ GeV$, respectively.}
\end{figure}

\begin{figure}[H]
    \centering
    \setlength{\abovecaptionskip}{-0.1cm}
    \includegraphics[width=7.5cm]{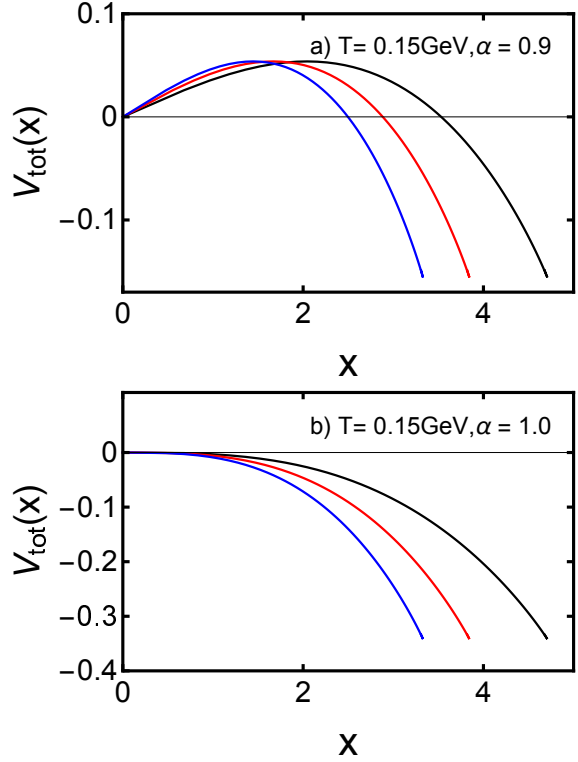}
    \caption{\label{fig9} The total potential $V_{tot}$ against the separate length x for the different magnetic fields when $T = 0.15\ GeV$. a) for $\alpha = 0.9$ and b) for $\alpha = 1.0$. The black line, red line and blue line in a) and b) denote $B = 0.1\ GeV^{2}$, $0.15\ GeV^{2}$ and $0.2\ GeV^{2}$ respectively.}
\end{figure}

In Fig.~\ref{fig8}, we plot $E_{c}$ against $B$ when $T = 0.15\ GeV$ and find that the $E_{c}$ increases with $B_{\perp}$ which is consistent with the results in \cite{Sato:2013pxa,Hashimoto:2014yya}, which is different from our result for the weak magnetic field shown in Fig.~\ref{fig4}. The reasons may due to the different ways of turning on the magnetic field. In this paper, the magnetic field affects the geometry of background and has an influence on the potential barrier. Moreover, we find the high temperature also reduces $E_{c}$  consistent with the finding in Fig.~\ref{fig5} for weak magnetic field case.

The effect of the magnetic field on the total potential when $T = 0.15\ GeV$ in different external electric-field is studied in Fig.~\ref{fig9}. When $\alpha = 0.9$, the magnetic field enhance the total potential in small distance $x$. However, the effect of the magnetic field on the width of the potential barrier is more prominent in large distance $x$. The magnetic field reduces the width of the potential barrier and enhance the Schwinger effect in large distance $x$ although the magnetic field enhances $E_{c}$. When $\alpha = 1.0$, the magnetic field reduces the width of the potential barrier obviously and favors the Schwinger effect.

It should be mentioned that the magnetic field has no effect on separate length and the sum of the Coulomb potential and static energy when the pairs are in parallel to the magnetic field. In this case, $E_{c}$ increases with magnetic field and Schwinger effect is suppressed.

\section{Conclusion and discussion}\label{sec:05}
In this paper, we study the potential analysis in the 5-dimensional Einstein-Maxwell system with the magnetic fields  corresponding  to the RHIC and LHC energies. Since the heavy ion collisions at RHIC and LHC experiments produce strong electro-magnetic fields. The strong magnetic fields may provide some different views for the vacuum structure and we expect that the Schwinger effect could be observed through the heavy-ion collisions  in future.

The separate length between test particle pairs by using a probe D3-brane at a finite radial position is discussed in this article. We consider the test particle pairs both transverse to the magnetic field and parallel to the magnetic field. We  find that the separating length decreases with the increasing magnetic field and the temperature.

We calculated the critical electric field via the DBI action and derived the formula of the total potential so that we can perform the potential analysis in the magnetized backgrounds. It is found that both the magnetic field and the temperature reduce the potential barrier and the critical field with the weak magnetic field  $B\ll T^2$ solutions, thus enhance the Schwinger effect. That means the magnetic field and the temperature increases the production rate of the real particle pairs. For the strong magnetic field case with $B\gg T^2$ solutions when the pairs are in perpendicular to the magnetic field, the magnetic field also enhances the Schwinger effect rate though the magnetic field increases the critical electric field since magnetic field reduces the width of the potential barrier and enhences potential at larger distance.

We expect that the nontrivial magnetic field effects on the Schwinger effect in the magnetized background could provide some inspiration of QCD with a strong electric field. Moreover, the production rate in the Einstein-Maxwell-dilaton system in a holographic QCD model may be worth to be investigated \cite{DeWolfe:2010he,Yang:2014bqa,Chen:2019rez,ChenXun:2019zjc}. We hope to report in these directions in future.

\section*{Acknowledgments}

This work is in part supported by the NSFC Grant Nos. 11735007, 11890711.

\end{document}